\documentclass[aps,showpacs]{revtex4}
\usepackage{amsmath}
\usepackage{graphicx}

\begin{document}

\title{Frenet algorithm for simulations of fluctuating continuous
elastic filaments}
\author{Yevgeny Kats}
\author{David A. Kessler}
\author{Yitzhak Rabin}
\affiliation{Dept. of Physics, Bar-Ilan University, Ramat-Gan, Israel}
\date{\today}

\begin{abstract}
We present a new algorithm for generating the equilibrium configurations of
fluctuating continuous elastic filaments, based on a combination of
statistical mechanics and differential geometry. We use this to calculate
the distribution function of the end-to-end distance of filaments with
nonvanishing spontaneous curvature and show that for small twist and large
bending rigidities, there is an intermediate temperature range in which the
filament becomes nearly completely stretched. We show that volume
interactions can be incorporated into our algorithm, demonstrating this
through the calculation of the effect
of excluded volume on the end-to-end distance of the filament.
\end{abstract}

\maketitle

Theories and computer simulations of polymers are based on the notion that a
macromolecule can be modeled as a collection of points with positions $%
\left\{ \mathbf{r}_{i}\right\} $ that represent either ``real'' chemically
bonded monomers interacting through semi-empirical potentials (e.g., in
Molecular Dynamics simulations\cite{Dennis}), or statistically independent
segments connected by elastic springs or subject to constraints (in Monte
Carlo simulations\cite{Binder}). In the latter case, the elastic energy is
usually assumed to be of entropic origin\cite{Flory} $E_{el}=\left(
K/2\right) \sum_{i}(\mathbf{r}_{i}-\mathbf{r}_{i-1})^{2},$ where $%
K=k_{B}T/aL $ is the spring constant, $k_{B}$ the Boltzmann constant, $T$
the temperature, $L$ the polymer length and $a$ the monomer size. In polymer
physics one often employs the continuum version of this theory, the 
so-called Gaussian chain (GC) model, 
in which the monomer label is replaced by a
continuous contour parameter $s$, $\mathbf{r}_{i}\rightarrow $ $\mathbf{r}%
(s) $ and $E_{GC}=$ $\left( K^{\prime }/2\right) \int_{0}^{L}ds\left( d%
\mathbf{r/}ds\right) ^{2}$ ($K^{\prime }$ is a constant)$.$ Alternatively,
one can use the worm-like chain (WLC) model in which the energy penalty for
stretching of elastic springs is replaced by bending elasticity, $%
E_{WLC}=\left( k_{B}Tl_{p}/2\right) \int_{0}^{L}ds\left( d^{2}\mathbf{r/}%
ds^{2}\right) ^{2}$ where $l_{p}$ is the bending persistence length. At
first sight it appears that in taking the continuum limit we pass from a
description of a polymer as a collection of points to one in which it is
described as a line. However, any geometrical line in 3d space must obey the
Pythagorean theorem, $\left| d\mathbf{r(}s)/ds\right| =1$, a condition that
can not be imposed in the framework of the GC model (it would make $E_{GC}$
a conformation-independent constant). This is consistent with the
observation that the statistical properties of the GC model are identical to
a those of a random walk and therefore the conformation of a polymer in this
model is a fractal, with fractal dimension 2 (recall that the fractal
dimension of a line is the same as its geometric dimension, 1). Although the
above constraint is consistent with $E_{WLC}$, the resulting nonlinear
theory appears to be intractable. Nevertheless, it was shown that the
statistical mechanics of this model can be worked out using the analogy
between the WLC and a quantum top, and the results were successfully applied
to model the stretching of DNA molecules\cite{MS94}.

Recently, we analyzed the statistical mechanics of the generalized WLC model
that describes the linear elasticity of ribbons, with elastic energy\cite%
{Helix} 
\begin{equation}
E_{el}=\dfrac{1}{2}\sum_{k=1}^{3}b_{k}\int_{0}^{L}ds\left(
\omega_{k}-\omega_{k0}\right) ^{2},  \label{energy}
\end{equation}
where the coefficients $b_{1}$ and $b_{2}$ are associated with the bending
rigidities with respect to the two principal axes of inertia of the cross
section (they differ if the cross section is not circular), and $b_{3}$
represents twist rigidity. In this paper we will treat $\left\{
b_{i}\right\} $ as given material parameters of the filament. The functions $%
\left\{ \omega_{k}(s)\right\} $ and $\left\{ \omega_{k0}(s)\right\} $ are
related to the generalized curvatures and torsions in the deformed and the
stress-free states of the filament, respectively. These parameters
completely determine the three dimensional conformation of the centerline
and the twist of the cross-section about this centerline, through the
generalized Frenet equations 
\begin{equation}
\dfrac{d\mathbf{t}_{i}(s)}{ds}=-\sum_{j,k}\varepsilon_{ijk}\omega _{j}(s)%
\mathbf{t}_{k}(s).  \label{Frenet}
\end{equation}
Here $\mathbf{t}_{3}$ is the unit tangent to the centerline and the unit
vectors $\mathbf{t}_{1}$ and $\mathbf{t}_{2}$ are oriented along the
principal axes of the cross section ($\varepsilon_{ijk}$ is the
antisymmetric tensor). Note that since these equations describe a pure
rotation of the triad $\left\{ \mathbf{t}_{i}(s)\right\} $ as one moves
along the contour of the filament, the constraint $\left| \mathbf{t}%
_{3}\right| =\left| d\mathbf{r(}s)/ds\right| =1$ is automatically satisfied
in this intrinsic coordinate description.

Since the energy is a quadratic form in the deviations $\delta\omega
_{k}=\omega_{k}-\omega_{k0}$ of the curvature and torsion parameters from
their values in the stress free state (Eq. (\ref{energy}) is valid as long
as the characteristic length scale of the deformation is larger than the
diameter of the filament\cite{Love}), the distribution of $\delta\omega_{k}$
is Gaussian, with zero mean and second moments given by\cite{Helix}

\begin{equation}
\left\langle \delta \omega _{i}(s)\delta \omega _{j}(s^{\prime
})\right\rangle =\dfrac{k_{B}T}{b_{i}}\delta _{ij}\delta (s-s^{\prime }).
\label{corr}
\end{equation}%
Using the above expression we showed that all the two-point correlation
functions $\left\langle \mathbf{t}_{i}(s)\cdot \mathbf{t}_{j}(s^{\prime
})\right\rangle $ can be calculated by solving a simple linear differential
equation with $s$-dependent coefficients, for arbitrary parameters of the
stress free state $\left\{ \omega _{k0}\right\} $ and rigidity parameters $%
\left\{ b_{k}\right\} $. However, since the distribution functions of the
various fluctuating quantities (e.g., the end-to-end distance $R$) are
non-Gaussian, knowledge of the second moments does not determine the
distributions and therefore, the complete determination of the statistical
properties of fluctuating ribbons requires more powerful analytical or
numerical methods. In this paper we present an efficient algorithm for the
simulation of fluctuating elastic ribbons and use it to study the effects of
spontaneous curvature and twist rigidity on the spatial conformations of
fluctuating ribbons. To the best of our knowledge this is the first direct
simulation of continuous lines (other simulations of the WLC model and its
extensions represent the filament as a collection of points \cite%
{Frey,Kremer,Maggs}).

Examination of Eq. (\ref{corr}) shows that the decoupling of the ``noises'' $%
\left\{ \delta \omega _{k}\right\} $ in the intrinsic coordinate
representation permits an efficient numerical generation of independent
samples drawn from the exact canonical distribution. The Gaussian
distribution of the $\delta \omega _{k}$ means that each $\delta \omega
_{k}(s)$ can be directly generated as a string of independent random numbers
drawn from a distribution symmetric about the origin with width $\sqrt{%
k_{B}T/b_{k}ds},$ where $ds$ is the discretization step length. Note that
the discretization of the continuous filament (choice of $ds$) is a
computational tool only, and is very different from the case of a chain
consisting of discrete mass points. We always choose $ds$ sufficiently small
so that the results are insensitive to its precise value.

The remaining task is to construct the curve using the Frenet equations with 
$\omega _{k}(s)=\omega _{k0}(s)+\delta \omega _{k}(s)$. The Frenet equations
are best integrated by stepping the basic triad $\left\{ \mathbf{t}%
_{k}\right\} $ forward in $s$ through a suitable small rotation. In this
way, the orthonormality of the triad is guaranteed to be preserved up to
machine accuracy. To construct this rotation matrix, we begin with Eq. (\ref%
{Frenet}) and, defining the three vectors $%
v^{x}=(t_{1}^{x},t_{2}^{x},t_{3}^{x})$, and so forth, we can write this
equation as 
\begin{equation}
\frac{dv^{i}}{ds}=Av^{i}  \label{Aeq}
\end{equation}%
where $A$ is an antisymmetric matrix with elements $A_{ij}=\sum_{k}%
\varepsilon _{ijk}\omega _{k}$. We now discretize Eq. (\ref{Aeq}) as 
\begin{equation}
v^{i}(s+ds)=Ov^{i}(s)
\end{equation}%
where the orthogonal matrix $O$ is 
\begin{equation}
O=\left( 1+\frac{ds}{2}A\right) \left( 1-\frac{ds}{2}A\right) ^{-1}
\end{equation}

In the following we present the results of simulations of distribution of
the end-to-end distance for a filament of contour length $L=1$ and study its
dependence on the stress free configuration $\left\{ \omega _{k0}\right\} ,$
rigidity parameters $\left\{ b_{k}\right\} $ and temperature.

We first consider a straight filament, with $\omega _{10}=\omega
_{20}=\omega _{30}=0$. In the limit $T\rightarrow 0$, the distribution $P(R)$
of the end-to-end distance, $R$, approaches a delta function peaked at $R=1$%
. With increasing $T,$ the peak of the distribution shifts to smaller values
of $R,$ consistent with the fact that the effective persistence length
scales as $l_{p}\sim b/k_{B}T$ ($b$ is a characteristic bending rigidity
parameter). The behavior of the width of the distribution is more
interesting. Initially (in the range $l_{p}\gg 1$) the distribution broadens
with $T$ and then narrows down again as the Gaussian chain limit $%
\left\langle \mathbf{R}^{2}\right\rangle =l_{p}L\propto 1/T$ is approached
for $l_{p}\ll 1$. This behavior is universal and takes place for arbitrary
values of the rigidity parameters; furthermore, the form of the distribution
depends only on the bending rigidities $b_{1},$ $b_{2}$ and is unaffected by
the twist rigidity $b_{3}.$

\begin{figure}
\includegraphics[width=3.5in,clip]{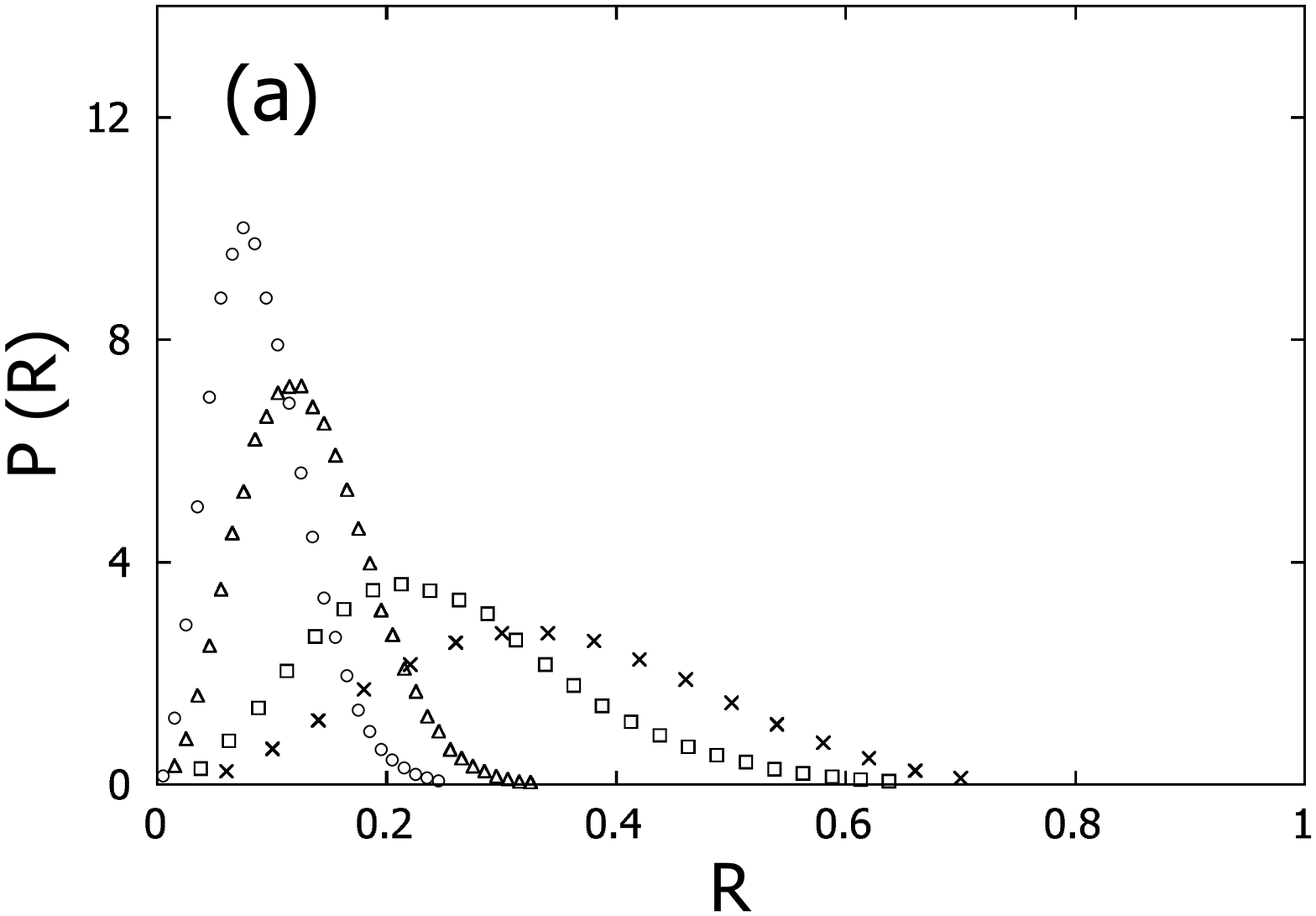} %
\includegraphics[width=3.5in,clip]{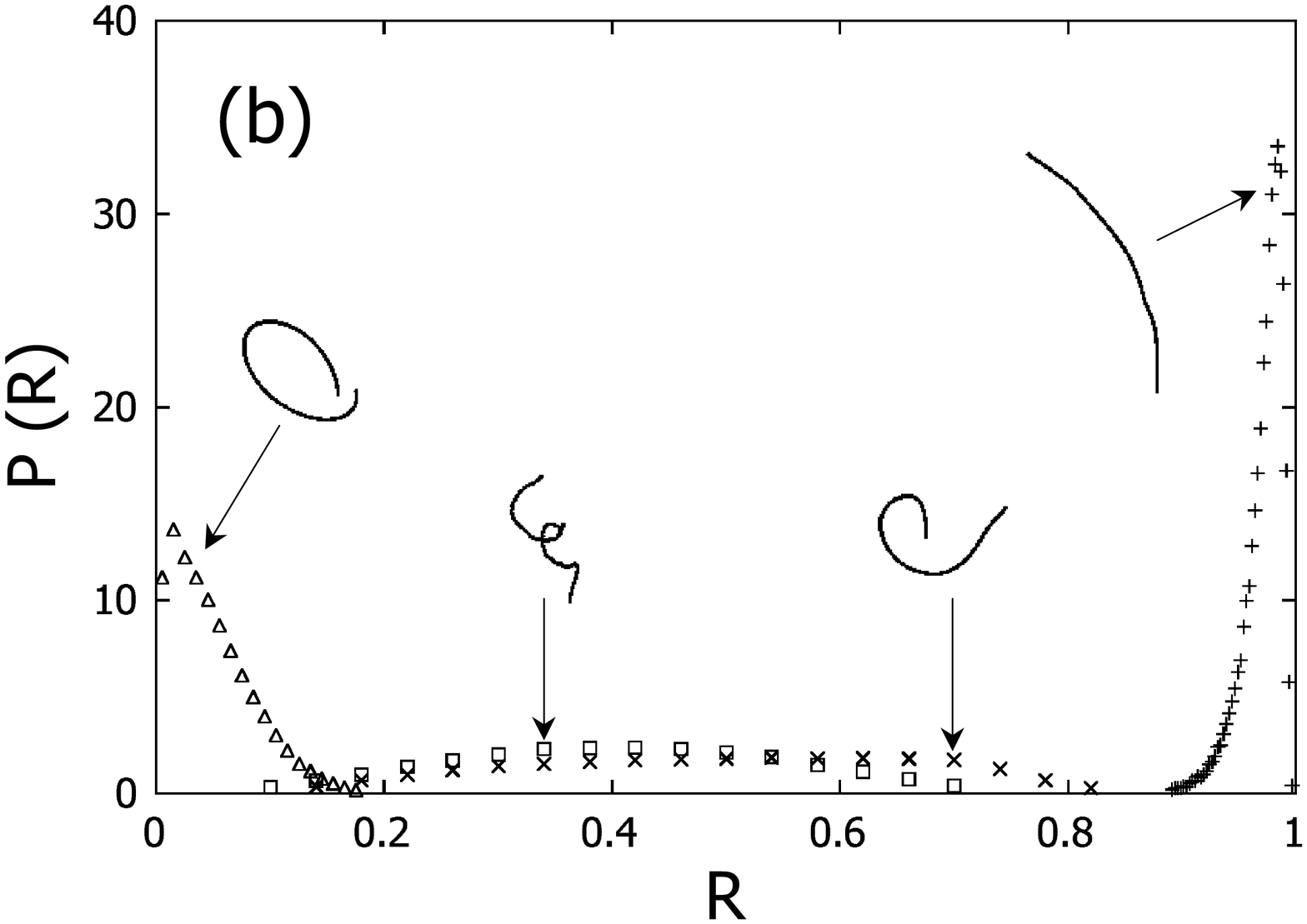}
\caption{Plot of the distribution function $P(R)$ vs. the end-to-end to end
distance $R$ for an open ring. a) $b_{1}=b_{2}=b_{3}=1$, and temperatures $%
T=0.1$ (circles), 1 (squares), 10 (crosses) and 100 (triangles). b) $%
b_{1}=b_{2}=1$, $b_{3}=10^{-4}$, and temperatures $T=10^{-5}$ (triangles), $%
10^{-3}$ (crosses), 0.1 (pluses) and 10 (squares). Snapshots of typical
configurations corresponding to each of these temperatures are shown as
inserts.}
\end{figure}

Now consider a ring, broken at a point so that the ends are free to
fluctuate. Here $\omega _{10}=2\pi $, $\omega _{20}=\omega _{30}=0$. In Fig.
1a we present the distribution function $P(R)$ for the case $%
b_{1}=b_{2}=b_{3}=1$. As dictated by the geometry of the stress free state, $%
P(R)$ approaches a delta function peaked at $R=0$ in the limit $T\rightarrow
0.$ At higher temperatures, fluctuations increase $R$, the distribution
broadens and its peak moves out to higher values of $R$ of the order of the
diameter of the ring ($1/\pi $). At yet higher $T$ the decrease of the
persistence length with increasing temperature takes over, the distribution
narrows and its peak moves to smaller values of $R$.

In Fig. 1b we present the case of the broken ring now at small twist
rigidity ($b_{3}\ll b_{1},b_{2}$). While the low temperature behavior is
similar to that in Fig 1a, nearly full stretching accompanied by a dramatic
narrowing of the distribution of the end-to-end distance is observed at
intermediate temperatures for which $a_{1}/L=a_{2}/L>1,$ $a_{3}/L\ll 1$ (we
define the bare persistence lengths, $\left\{ a_{i}\right\} =\left\{
b_{i}/k_{B}T\right\} $). This striking observation is supported by
analytical results for the mean square end-to-end distance\cite{Helix} $%
\left\langle \mathbf{R}^{2}\right\rangle
=2\int_{0}^{L}ds\int_{0}^{s}ds^{\prime }\left[ e^{-\mathbf{\Lambda }\left(
s-s^{\prime }\right) }\right] _{33},$ where the elements of the matrix $%
\mathbf{\Lambda }$ are $\Lambda _{ik}=\left( k_{B}T/2\right) \left[
\sum_{j}1/b_{j}-1/b_{i}\right] \delta _{ik}+\sum_{l}\varepsilon _{ikl}\omega
_{l0}.$ An exact expression for $\left\langle \mathbf{R}^{2}\right\rangle $
in terms of the parameters $\left\{ b_{k}\right\} $ was derived by
diagonalizing the matrix $\mathbf{\Lambda }$ and is compared against our
simulation results in Fig. 2. We would like to emphasize that the nearly
complete stretching of the filament is a finite $L$ effect and that the
renormalized persistence length defined by $l_{p}=\lim_{L\rightarrow \infty
}\left\langle \mathbf{R}^{2}\right\rangle /L,$ does not diverge in the limit
of vanishing twist rigidity.

\begin{figure}[tbp]
\includegraphics[width=4.in,clip]{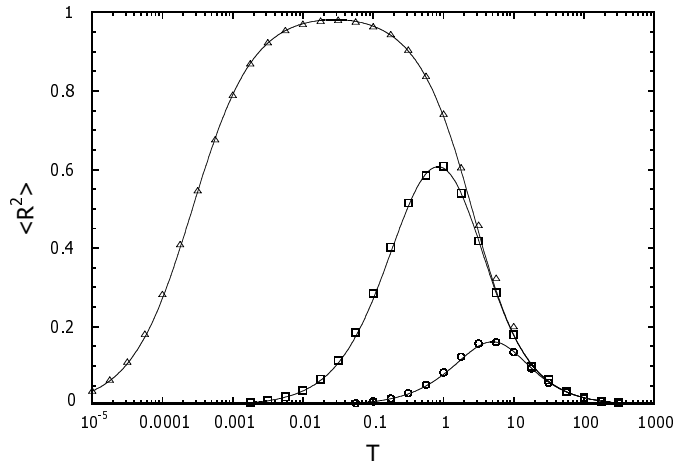}
\caption{Simulation results for the mean square end-to-end distance $%
\left\langle R^{2}\right\rangle $ vs. temperature $T$ for an open ring, with 
$b_{1}=b_{2}=1$ and $b_{3}=1$ (circles), $b_{3}=10^{-2}$ (squares) and $%
b_{3}=10^{-5}$ (triangles). The results of the analytical calculations are
shown as solid lines.}
\end{figure}

The observation of nearly complete stretching of the filament is at first
sight counterintuitive, given that at these temperatures, its bending
persistence length is larger than its contour length, and so thermal
fluctuations can not significantly change the curvature from its finite
spontaneous value ($\omega _{10}=2\pi $), let alone cause it to vanish. The
resolution of this paradox is that our intuition that a circle can not be
continuously deformed into a line, at fixed curvature, is wrong. Any 3d
curve can be described by its curvature $\omega _{1}$ and torsion $\omega
_{3}$ (if we replace the filament by a geometrical line, we should set\cite%
{Helix} $\omega _{2}=0$). Keeping the curvature fixed and increasing the
torsion (assumed to be constant along the filament), one goes from a circle
of radius $\omega _{1}^{-1}$ to an increasingly stretched helix, with radius 
$\omega _{1}/\left( \omega _{1}^{2}+\omega _{3}^{2}\right) \rightarrow 0$
and end-to-end separation approaching that of a straight line. This indeed
takes place in our case since Eq. (\ref{corr}) gives $\left\langle \omega
_{3}^{2}\right\rangle =k_{B}T/(b_{3}ds)$ so that thermal fluctuations of the
torsion $\omega _{3}$ are large in the limit of small twist rigidity. Since,
under conditions of nearly complete stretching, the distribution of the
end-to-end distance is extremely narrow (see Fig. 1b), mean field
considerations apply and we can describe the filament as an object with
length $L=1,$ curvature $\omega _{1}=2\pi $ and torsion $\left\langle \omega
_{3}^{2}\right\rangle ^{1/2}=\pm \sqrt{k_{B}T/(b_{3}ds)}.$ For high
temperature and small twist rigidity, $|\left\langle \omega
_{3}^{2}\right\rangle ^{1/2}|\gg \omega _{10}$, this object is a nearly
straight helix that oscillates between left-hand and right-hand forms as one
moves along its axis (in this limit the change of sign of torsion changes
the sense of helical rotation but does not affect the direction of the axis
about which the helix rotates), with typical radius of helical turn $2\pi
b_{3}ds/(k_{B}T)$ that is much smaller than its pitch $2\pi \sqrt{%
b_{3}ds/(k_{B}T)}.$ Since in this limit the pitch approaches the period of
the helix, we conclude that the helical filament is nearly completely
stretched. At yet higher temperatures, the bending persistence length
becomes shorter than the length of the filament, the helical structure
``melts'', and one recovers the Gaussian chain behavior characteristic of
flexible polymers.

In order to demonstrate that volume interactions between parts of the
filament can be readily incorporated into our algorithm, we calculated the
Flory exponent $\nu $ defined by the relation, $\left\langle \mathbf{R}%
^{2}\right\rangle \sim L^{2\nu }$. We inserted $50$ equally spaced
interaction sites per unit length into the filament ($50L$ sites in total)
and allowed only configurations in which the spatial distance between any
two of these sites exceeded $0.01.$ For each value of $L$ we generated $1000$
allowed configurations and calculated the average value of $\mathbf{R}^{2}$.
We first checked that when all configurations were allowed (no excluded
volume), we obtain the Gaussian random walk result, $\nu _{0}\approx0.51.$ 
In the
presence of excluded volume, the best fit to the simulation (Fig. 3), yields 
$\nu _{SAW}\approx0.59,$ in excellent agreement with the Flory exponent of $3/5$
for self-avoiding polymers in 3d\cite{Flory}. 
\begin{figure}[tbp]
\includegraphics[width=4.in,clip]{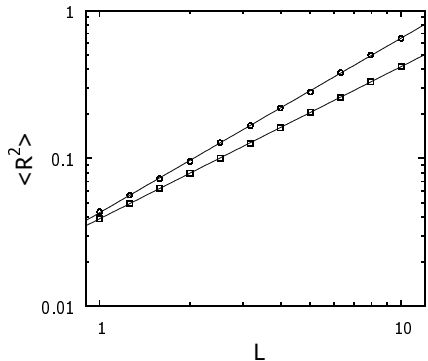}
\caption{Simulation results for the mean square end-to-end distance $%
\left\langle R^{2}\right\rangle $ vs. the contour length $L$ for a linear
filament, with (circles) and without (squares) excluded volume. The solid
lines give the least square fit to the data. The parameters are $%
b_{1}=b_{2}=b_{3}=0.02$, and $T=1.$}
\end{figure}

In closing, we would like to comment on possible ramifications of this work.
The methods presented in this work are ideally suited to investigate
polymers with spontaneous curvature such as double stranded DNA\cite{TTH87},
and synthetic molecules with bent cores\cite{Link}. As far as equilibrium
properties are concerned, our algorithm is computationally far superior to
standard Monte-Carlo methods in that each realization of the filament is
completely independent. Because sequence dependence can be readily
incorporated into the present theory through the dependence of the
spontaneous curvature and torsion parameters $\left\{ \omega
_{k0}(s)\right\} $ on the contour position $s,$ and because the method is
easily extendible to incorporate excluded volume, electrostatic and other
self-interaction effects, the Frenet algorithm has the potential of becoming
an important new tool for computer simulations of equilibrium properties of
complex biopolymers such as DNA, RNA and proteins.

\begin{acknowledgments}
The assistance of Merav Eshed in the numerical computations is gratefully
acknowledged. DAK and YR acknowledge the support of the Israel Science
Foundation.  YR thanks K. Binder for useful comments on a previous version
of the manuscript.  DAK thanks J. Schiff for a useful discussion on the
integration procedure.
\end{acknowledgments}

\newpage \printfigures

\end{document}